\documentclass[preprint,aps,prc,showpacs,superscriptaddress,longtitle]{revtex4-2}

\usepackage{graphicx}
\usepackage{amsmath}
\usepackage{amssymb}
\usepackage{amsfonts}
\usepackage{bm}
\usepackage{braket}
\usepackage{hyperref}
\usepackage{color}
\usepackage{float}
\usepackage{subcaption}
\usepackage{epstopdf}
\usepackage{booktabs}

\begin{document}

\title{Effects of Strong Magnetic Fields on the Equation of State and Mass-Radius Structure of Hyperonic Neutron-Star Matter with Anomalous Magnetic Moments}

\author{Wazha German}
\author{Jacobus P. W. Diener}
\author{Tumisang Mello}
\author{Gregory Hillhouse}
\affiliation{Department of Physics and Astronomy, Botswana International University of Science and Technology, Palapye, Botswana}

\begin{abstract}
\thispagestyle{empty}
We investigate the effects of strong magnetic fields on the equation of state (EoS) and stellar structure of cold, charge-neutral, $\beta$-equilibrated hyperonic neutron-star matter within a relativistic mean-field (RMF) approach. We include leptons and the full baryon octet. All charged particles are Landau quantized, and all baryons, including the neutral ones, couple to the magnetic field through their anomalous magnetic moments (AMM). The calculation is performed with the FSU2H parameterization, with FSU2R used to test the parameter dependence, and is compared for zero field, a constant magnetic field, and density-dependent magnetic-field configurations. We find that, at fixed baryon density, strong magnetic fields modify the hyperonic composition through the competing effects of Landau quantization and AMM-induced spin splitting, with small, species-dependent changes in the particle fractions. At fixed total energy density, the inclusion of AMM stiffens the global EoS over most of the energy-density range, typically by about $0.1\%$, albeit with narrow windows of local softening. The associated increase in the maximum supported neutron-star mass depends on the magnetic-field configuration and is only about $0.002\,M_\odot$. The present work provides a benchmark for assessing the influence of AMM on the composition, magnetization, equation of state, and mass--radius structure of magnetized hyperonic neutron-star matter.
\end{abstract}

\maketitle

\newpage

\section{INTRODUCTION}\label{sec:introduction}
We use the hyperonic equation of state of Tolos et al.~\cite{Tolos2017} to define the underlying relativistic mean-field model. Neutron stars are among the most extreme astrophysical objects in the Universe, serving as natural laboratories for strongly interacting matter under conditions of density, isospin asymmetry, and magnetic field strength that cannot be replicated in terrestrial experiments \cite{Glendenning2000,Lattimer2007,Haensel2007}. Their central densities can reach several times nuclear saturation density \(n_0 \approx 0.15\, \mathrm{fm}^{-3}\), corresponding to pressures and energy densities that probe the fundamental interactions between hadrons. The internal composition and equation of state (EoS) of neutron star matter remain central open problems in nuclear astrophysics, with direct implications for our understanding of the strong interaction in the non-perturbative regime.

These questions become particularly compelling in the context of magnetars, a subclass of neutron stars characterized by surface magnetic fields of order \(10^{14} - 10^{15}\) G \cite{Vasisht1997,Kouveliotou1998,Woods1999,Mereghetti2008,Rea2011,Turolla2015}. In the interior, field strengths may reach \(10^{17} - 10^{18}\) G or higher, far exceeding the critical quantum electrodynamic field \(B_{\mathrm{crit}} = m_e^2 c^3 / (e\hbar) \approx 4.4 \times 10^{13}\) G \cite{Broderick2000}. The origin of such extreme fields remains debated, with proposed mechanisms including flux conservation during core collapse, dynamo action in rapidly rotating proto-neutron stars, and magneto-rotational instabilities (for a recent review, see Ref.~\cite{Sinha2026}). Such objects provide a unique setting in which strong-interaction physics, relativistic many-body theory, and quantum electrodynamic effects must be considered simultaneously.

The presence of strong magnetic fields modifies dense matter in several fundamental ways. Charged particles undergo Landau quantization, wherein their transverse momentum is discretized into distinct Landau levels, fundamentally altering the density of states and thus the particle populations and thermodynamic properties of the system \cite{Broderick2000,Broderick2002,Cardall2001}. Additionally, the magnetic field contributes directly to the energy density and pressure, potentially affecting the stiffness of the EoS and the macroscopic properties of compact stars \cite{Rabhi2010,Lopes2012,Gomes2014}. The impact of strong magnetic fields on neutron star matter has been investigated in various contexts, including nucleonic matter, quark matter, hybrid matter, and hyperonic matter \cite{Chakrabarty1997,Ferrer2005,Ferrer2006,Sinha2013,Tolos2017}. Beyond hyperons, strong magnetic fields may also affect the onset of \(\Delta\) resonances, meson condensates, and deconfined quark matter \cite{Sinha2026}. 

We investigate whether the anomalous magnetic moments (AMM) of baryons materially affect the equation of state of magnetized hyperonic neutron-star matter. To address this, we take the FSU2H parameter set within relativistic mean-field (RMF) approach of Tolos et al.~\cite{Tolos2017}, add full baryon-octet AMM, and systematically compare results with and without AMM. While the present work focuses exclusively on the baryon octet with AMM, the interplay between multiple exotic degrees of freedom in magnetized matter remains an important open problem.

Throughout this work, we distinguish the scalar thermodynamic matter pressure from the directional components of the stress tensor. Magnetization-current forces must be included when interpreting the mechanical response of magnetized matter \cite{Blandford1982,Potekhin2012,Chatterjee2015}. This distinction does not remove the directional Maxwell stresses of an ordered magnetic field. The pressure prescription adopted for the stellar-structure calculations is specified in Sec.~II.

Strong magnetic fields also affect neutrino emission processes in proto-neutron stars. Maruyama et al.~\cite{Maruyama2014} have demonstrated that neutrino production through the direct Urca process becomes asymmetric in strongly magnetized matter, with emission enhanced parallel to the magnetic field direction and suppressed in the opposite direction. The production asymmetry was found to have the same sign and comparable magnitude as previously established absorption and scattering asymmetries, implying that the net neutrino flux asymmetry is significantly enhanced when all three processes are considered. The particle fractions that determine Urca emissivity are themselves modified by magnetic fields and, as we show in this work, by AMM effects, creating a coupling between the microphysical EoS and macrophysical observables that warrants systematic investigation.

At densities exceeding approximately two to three times nuclear saturation density, the neutron chemical potential becomes sufficiently large to render the appearance of hyperons energetically favorable \cite{Glendenning2000,Haensel2007,Tolos2017,Chatterjee2016}. These strange baryons introduce additional degrees of freedom that generally soften the EoS, reducing the maximum mass that can be supported against gravitational collapse. This softening gives rise to the so-called hyperon puzzle: many hyperonic equations of state predict maximum masses below the observed \(\sim 2M_{\odot}\) neutron stars, such as PSR J1614-2230 \cite{Demorest2010}, whose refined mass is $1.937\pm0.014\,M_{\odot}$ and PSR J0740+6620, with $1.99\pm0.07\,M_{\odot}$ \cite{Agazie2023}. 
The earlier mass estimate for PSR J0348+0432 has also been revised from about $2.01\,M_{\odot}$ to $1.81\pm0.04\,M_{\odot}$ \cite{Antoniadis2013,Saffer2025}. 
Resolving this tension requires identifying mechanisms that can counteract hyperon softening while remaining consistent with nuclear and astrophysical constraints.

We follow the treatment of Landau quantization and baryon AMM developed by Broderick et al.~\cite{Broderick2000,Broderick2002}. The AMM term $-\frac12\kappa_b\bar\Psi_b\sigma_{\mu\nu}F^{\mu\nu}\Psi_b$ introduces spin-dependent shifts in the charged and neutral baryon spectra. Their effects depend on the field strength and composition and need not have the same sign for every species.

Yue et al.~\cite{Yue2009} studied hyperonic matter with Landau quantization and baryon AMM in an RMF model. Their displayed matter EoS excludes the free-field energy and pressure contributions. This distinction matters when comparing their results with a total EoS that includes electromagnetic stresses.

Tolos et al.~\cite{Tolos2017} introduced the FSU2R and FSU2H parameterizations based on FSU2~\cite{Chen2014}. FSU2R was calibrated to nucleonic stars, whereas FSU2H provides additional high-density stiffness to support massive stars containing hyperons. Their magnetized calculation omits baryon AMM: they argued that appreciable AMM effects occur at fields above those adopted in their study. The original parameters used here are distinguished from the subsequent isovector refit~\cite{TolosPASA2017} in Sec.~\ref{sec:rmf_sets}.

Recent work further clarifies the position of the present study within the broader literature. Sanson et al. \cite{Sanson2026} performed a systematic Bayesian comparison of density-dependent and nonlinear RMF models extended to the full baryon octet, emphasizing the sensitivity of hyperon onsets, particle fractions, and maximum masses to the poorly constrained hyperon couplings. Rather et al. \cite{Rather2021} studied heavy magnetic neutron stars in density-dependent RMF models and found that strong internal magnetic fields can induce re-leptonization and de-hyperonization, thereby stiffening hyperonic EoS. Wu et al. \cite{Wu2017} investigated hyperonic neutron stars using an FSUGold-based RMF model with chaotic magnetic fields and AMM, finding that magnetic fields stiffen the EoS while AMM can alter microscopic magnetic response and particle polarization. The finite-temperature extension of the Tolos et al. framework \cite{Tolos2017} has been investigated by Kochankovski et al.~\cite{Kochankovski2022}. The present work complements these studies by combining the FSU2 hyperonic RMF framework with Landau quantization, full baryon-octet AMM, magnetization, particle fractions, EoS, and TOV mass-radius calculations.

Including the AMM coupling allows neutral baryons to also interact directly with the magnetic field \cite{Broderick2000,Cardall2001,Diener2012,Yue2009}. Because the matter considered here is neutron rich, the magnetic response of neutral baryons is relevant. Indeed, previous studies have shown that AMM may compete with Landau quantization, influencing both the composition and the stiffness of the EoS in nontrivial ways \cite{Broderick2000,Yue2009,Dong2013}. While Landau quantization modifies the charged-particle density of states, the spin splitting introduced by the anomalous magnetic interaction modifies both charged and neutral baryon spectra. In the present calculation, this AMM contribution modifies the particle fractions and produces a small, energy-dependent change in pressure relative to the corresponding no-AMM hyperonic case.

The constant AMM couplings employed here are a phenomenological approximation. Schwinger treats magnetic-moment corrections and strong magnetic fields~\cite{Schwinger1998}; strong-field radiative calculations caution against extrapolating the weak-field lepton AMM correction linearly~\cite{ManrezaParet2014}. The relevant field scale depends on the particle mass:
\begin{equation}
B_{\mathrm{crit},b}
=
\frac{m_b^2c^3}{e\hbar}
=
4.414\times10^{13}
\left(\frac{m_b}{m_e}\right)^2
\,\mathrm{G}.
\end{equation}
For nucleons, this scale is approximately $1.5\times10^{20}$ G, so the fields of order $10^{18}$ G considered here remain below the scale set by the vacuum baryon masses. Consequently, the strong-field radiative results do not by themselves imply that the baryon AMM effects calculated here are overestimated. Electrons, in contrast, are far above their own critical field at these field strengths, and their weak-field AMM correction should not be extrapolated linearly. We neglect lepton AMM as a modelling assumption; failure of the weak-field extrapolation alone does not establish that every strong-field radiative correction is negligible. The baryon couplings used here are phenomenological values matched to vacuum magnetic moments \cite{Broderick2000,Yue2009,Dong2013}; possible in-medium modifications remain a separate source of uncertainty.

A separate but equally important consideration concerns the values of the AMM couplings themselves. Throughout this work and in most of the existing literature, vacuum magnetic moments are employed. However, in-medium modifications of baryon magnetic moments at finite density and temperature may be significant \cite{Ryu2010a}. Using the modified quark-meson coupling (MQMC) model, Ryu et al. \cite{Ryu2010b} have found that density-dependent AMMs lead to enhanced nucleon magnetic moments relative to hyperon moments at high density. In their treatment, this results in enhanced proton fractions and further suppression of hyperons. This highlights a significant model dependence: the competition between Landau quantization and AMM effects on hyperon populations and EoS stiffness may change depending on how AMMs evolve with density. A fully self-consistent treatment of density-dependent AMMs within the RMF framework remains an important direction for future work.

Mindful of these open questions, the present work explores the consequences of AMM within the standard phenomenological prescription using vacuum AMM values, with the understanding that our quantitative results represent one limiting case in an ongoing theoretical discussion. Pending the development of self-consistent density-dependent AMM treatments, our findings should be interpreted as representing AMM effects within a phenomenological RMF framework.

This work extends the framework of Broderick et al. \cite{Broderick2000} with Tolos et al.'s \cite{Tolos2017} updated FSU2H parameter set using both constant and density-dependent magnetic-field configurations, including the free electromagnetic field contributions to the EoS. This allows for a systematic comparison between constant and density-dependent magnetic-field configurations, with particular attention to particle fractions, magnetization, EoS stiffness, and the mass-radius relation.

This paper is organized as follows. In Sec. II we present the covariant RMF formalism used to describe magnetized hyperonic matter, including the anomalous magnetic coupling of baryons. In Sec. III we summarize the parameter sets and coupling schemes adopted in this work, including the hyperonic FSU2 parameterizations and the density-dependent magnetic-field profile. In Sec. IV we present our numerical results for the particle composition, Landau-level structure, magnetization, equation of state, and mass-radius relations. In Sec. V we discuss the implications of our findings, and in Sec. VI we present our conclusions.

\section{FORMALISM}

We follow the formalism of Refs.~\cite{Broderick2000,Broderick2002,Tolos2017,Yue2009} to describe dense hyperonic neutron-star matter in a strong magnetic field within the RMF approximation \cite{Serot1986,Serot1997}. The matter consists of the full baryon octet,
\begin{equation}
    b = \{p,n,\Lambda,\Sigma^{+},\Sigma^{0},\Sigma^{-},\Xi^{0},\Xi^{-}\},
    \label{eq:baryon_octet}
\end{equation}
together with electrons and muons, satisfying charge neutrality and $\beta$-equilibrium.

\subsection{Baryonic Lagrangian and AMM coupling}

The baryonic Lagrangian density is \cite{Broderick2000,Tolos2017}
\begin{multline}
\mathcal{L}_b = \sum_b \bar{\Psi}_b \Big[ i\gamma^\mu \partial_\mu - q_b \gamma^\mu A_\mu - m_b + g_{\sigma b} \sigma \\
- g_{\omega b} \gamma^\mu \omega_\mu - g_{\rho b} \gamma^\mu \mathbf{I}_b \cdot \boldsymbol{\rho}_\mu - g_{\phi b} \gamma^\mu \phi_\mu - \frac{1}{2} \kappa_b \sigma_{\mu\nu} F^{\mu\nu} \Big] \Psi_b.
\label{eq:baryon_lagrangian}
\end{multline}

The AMM term couples the anomalous part of the baryon magnetic dipole moment to the field \cite{Broderick2000,Broderick2002}. Writing the electric charge as $q_b=eQ_b$, where $Q_b$ is the dimensionless charge number, the dimensional AMM coupling is
\begin{equation}
\kappa_b
=
\left(
\frac{\mu_b^{\mathrm{mag}}}{\mu_N}
-
Q_b\frac{m_N}{m_b}
\right)\mu_N,
\qquad
\mu_N=\frac{e}{2m_N},
\label{eq:amm_definition}
\end{equation}
in units with $\hbar=c=1$. Here $\mu_b^{\mathrm{mag}}$ denotes the total magnetic moment, distinguished from the chemical potential $\mu_b$, and $\mu_N$ is the nuclear magneton. For neutral baryons, $Q_b=0$ and $\kappa_b=\mu_b^{\mathrm{mag}}$. The magnetic moments used to determine these couplings are listed in Table~\ref{tab:amm_values}. The $\Sigma^0$ value is estimated using the isospin relation \cite{Tsushima2022}
\begin{equation}
\mu_{\Sigma^0}^{\mathrm{mag}}=(\mu_{\Sigma^+}^{\mathrm{mag}}+\mu_{\Sigma^-}^{\mathrm{mag}})/2,
\end{equation}
while the other entries are the adopted rounded magnetic moments \cite{PDG2020}. For charged baryons, the Dirac contribution is subtracted according to Eq.~(\ref{eq:amm_definition}).

\begin{table}[ht]
\caption{Total baryon magnetic moments used to determine the AMM couplings. Values are expressed as multiples of the nuclear magneton $\mu_N$. }
\begin{tabular}{lrrrrrrrr}
\toprule
Baryon & $p$ & $n$ & $\Lambda$ & $\Sigma^+$ & $\Sigma^0$ & $\Sigma^-$ & $\Xi^0$ & $\Xi^-$ \\
\midrule
$\mu_b^{\mathrm{mag}}/\mu_N$
& 2.793 & -1.913 & -0.613 & 2.458 & 0.649 & -1.160 & -1.250 & -0.651 \\
\bottomrule
\end{tabular}
\label{tab:amm_values}
\end{table}

\subsection{Mean-field approximation and single-particle spectra}
In the RMF approximation, meson fields are replaced by their expectation values \cite{Serot1986,Serot1997}. The baryon effective mass is $m_b^* = m_b - g_{\sigma b}\bar{\sigma}$, and the vector shift is $V_b = g_{\omega b}\bar{\omega} + g_{\rho b}I_3^b\bar{\rho} + g_{\phi b}\bar{\phi}$.
For a magnetic field pointing along the $z$-axis,  $\bm{B}=B\hat{z}$, the single-particle spectra of charged baryons with AMM are \cite{Broderick2000,Broderick2002}:
\begin{equation}
E_{b,\nu,s} = \sqrt{k_z^2 + \left( \sqrt{m_b^{*2} + 2\nu |q_b| B} - s \kappa_b B \right)^2 } + V_b,
\label{eq:charged_spectrum}
\end{equation}
where $s=\pm1$ is the spin projection and $\nu$ is the Landau level index. The allowed spin--Landau combinations satisfy
\begin{equation}
\nu
=
n+\frac{1}{2}
-\frac{s}{2}\operatorname{sgn}(q_b),
\qquad
n=0,1,2,\ldots,
\label{eq:landau_spin_relation}
\end{equation}
Consequently, the lowest Landau level, $\nu=0$, admits only one spin projection.

For neutral baryons:
\begin{equation}
E_{b,s} = \sqrt{k_z^2 + \left( \sqrt{k_\perp^2 + m_b^{*2}} - s \kappa_b B \right)^2 } + V_b.
\label{eq:neutral_spectrum}
\end{equation}

For leptons, whose AMM couplings are neglected in this calculation, the spectrum is
\begin{equation}
E_{\ell,\nu}
=
\sqrt{k_z^2+m_\ell^2+2\nu|q_\ell|B}.
\label{eq:lepton_spectrum}
\end{equation}

\subsection{Meson field equations and equilibrium conditions}

The meson field equations, chemical potentials, $\beta$-equilibrium, and charge neutrality follow the standard RMF procedure \cite{Broderick2000,Tolos2017,Serot1997}. The key constraints are:
\begin{eqnarray}
\mu_i = b_i\mu_n-Q_i\mu_e,\qquad
\sum_i Q_i n_i=0,\qquad
n_B=\sum_b n_b.
\label{eq:equilibrium}
\end{eqnarray}
Here $b_i$ is the baryon number and $Q_i=q_i/e$ is the electric charge number.

\subsection{Energy density, pressure, and magnetization}

The matter energy density is \cite{Broderick2000,Tolos2017}
\begin{equation}
\begin{aligned}
\epsilon_{\mathrm{matt}}
={}&
\sum_b \epsilon_b+\sum_\ell \epsilon_\ell
+\frac{1}{2}m_\sigma^2\bar{\sigma}^2
+\frac{1}{2}m_\omega^2\bar{\omega}^2
\\
&+
\frac{1}{2}m_\rho^2\bar{\rho}^2
+\frac{1}{2}m_\phi^2\bar{\phi}^2
+\frac{\kappa}{3!}(g_{\sigma N}\bar{\sigma})^3
\\
&+
\frac{\lambda}{4!}(g_{\sigma N}\bar{\sigma})^4
+\frac{\zeta}{8}(g_{\omega N}\bar{\omega})^4
\\
&+
3\Lambda_\omega
(g_{\rho N}g_{\omega N}\bar{\rho}\bar{\omega})^2.
\end{aligned}
\label{eq:energy_density}
\end{equation}
The scalar self-interaction coefficient $\kappa$ in this expression is distinct from the baryon AMM coefficients $\kappa_b$.

At zero temperature, the scalar thermodynamic matter pressure is
\begin{equation}
P_{\mathrm{matt}}
=
\sum_i\mu_i n_i-\epsilon_{\mathrm{matt}},
\label{eq:matter_pressure}
\end{equation}
where the sum includes baryons and leptons. The magnetization is defined by \cite{Blandford1982}
\begin{equation}
M
=
-\left(
\frac{\partial\Omega_{\mathrm{matt}}}{\partial B}
\right)_{\{\mu_i\}},
\label{eq:magnetization_definition}
\end{equation}
where $\Omega_{\mathrm{matt}}$ is the matter grand-potential density. Equivalently, for charge-neutral, $\beta$-equilibrated matter at zero temperature,
\begin{equation}
M
=
-\left(
\frac{\partial\epsilon_{\mathrm{matt}}}{\partial B}
\right)_{n_B}.
\label{eq:magnetization_fixed_density}
\end{equation}

The electromagnetic expressions below are written in Gaussian units. The magnetic energy density and total energy density are
\begin{equation}
\epsilon_B=\frac{B^2}{8\pi},
\qquad
\epsilon=\epsilon_{\mathrm{matt}}+\epsilon_B.
\label{eq:total_energy_density}
\end{equation}
When $B$ is expressed in gauss, $\epsilon_B$ is obtained in $\mathrm{erg\,cm^{-3}}$. Before combining it with matter quantities expressed in $\mathrm{fm}^{-4}$, we use the conversion factor of 
\begin{equation}
1\,\mathrm{fm}^{-4}
=
3.1615\times10^{35}\,\mathrm{erg\,cm^{-3}}.
\label{eq:energy_density_conversion}
\end{equation}
Likewise, the magnetic field entering the single-particle spectra is converted consistently to natural units; the numerical value of $B$ in gauss is not inserted directly into those spectra.

The scalar thermodynamic pressure must be distinguished from directional stress components. Before including the mechanical contribution of magnetization-current forces, the kinetic matter stresses are commonly expressed as $P_{\mathrm{matt}}$ parallel to the field and $P_{\mathrm{matt}}-MB$ perpendicular to it. Magnetization-current forces compensate this apparent anisotropy of the matter contribution \cite{Blandford1982,Potekhin2012,Chatterjee2015}. The electromagnetic field nevertheless contributes the directional Maxwell stress
\begin{equation}
T_{ij}^{\mathrm{Maxwell}}
=
\frac{B^2}{8\pi}\delta_{ij}
-
\frac{B_iB_j}{4\pi}.
\label{eq:maxwell_stress}
\end{equation}
For an ordered magnetic field pointing along the $+z$-axis, its longitudinal and transverse diagonal components are $-\epsilon_B$ and $+\epsilon_B$, respectively. For stellar-structure calculations, we use the chaotic-field pressure prescription \cite{Tolos2017,Wu2017}. An isotropic average of the Maxwell stress, with $\langle B_iB_j\rangle=B^2\delta_{ij}/3$, gives
\begin{equation}
P_{\mathrm{chaotic}}
=
P_{\mathrm{matt}}+\frac{\epsilon_B}{3}
=
P_{\mathrm{matt}}+\frac{B^2}{24\pi}.
\label{eq:chaotic_pressure}
\end{equation}
This is an assumed isotropic description of the magnetic stress for the spherical stellar calculation, rather than the stress of a globally uniform ordered field.

\subsection{Landau-level filling}

For a given spin projection $s$, the highest occupied Landau level in the field regime considered here is
\begin{equation}
\nu_{\max}^{b,s}
=
\left\lfloor
\frac{(E_{F,b}+s\kappa_b B)^2-m_b^{*2}}
{2|q_b|B}
\right\rfloor,
\label{eq:nu_max_spin}
\end{equation}
where $E_{F,b}=\mu_b-V_b$ is the effective Fermi energy. Only allowed spin--Landau states with real longitudinal Fermi momentum contribute. For leptons, this reduces to
\begin{equation}
\nu_{\max}^{\ell}
=
\left\lfloor
\frac{\mu_\ell^2-m_\ell^2}
{2|q_\ell|B}
\right\rfloor.
\label{eq:nu_max_lepton}
\end{equation}

\section{PARAMETER SETS AND MODEL SETUP}

\subsection{RMF parameter sets}\label{sec:rmf_sets}

We use both FSU2H and FSU2R, adopting the original parameter sets in Table~I of Tolos et al.~\cite{Tolos2017}. FSU2H supplies the main hyperonic results; extending FSU2R to the same baryon content provides a test of parameter dependence. FSU2R was designed for compact nucleonic stars. FSU2H modifies its isoscalar interactions to recover sufficient high-density stiffness when hyperons are included. In particular, the smaller $\omega$ self-coupling, $\zeta=0.008$ rather than $0.024$, reduces the suppression of vector repulsion at high density. The common masses and the differing scalar and vector couplings are listed in Table~\ref{tab:fsu2_params}. With the full baryon octet at zero field, the present maximum masses are approximately $2.025M_\odot$ and $1.768M_\odot$, respectively, close to the original published values $2.03M_\odot$ and $1.77M_\odot$.

The parameter version is important. A later publication by the same authors~\cite{TolosPASA2017} explicitly refits the isovector sector to improve the low-density neutron-matter behaviour. Those later sets are distinct from the original ApJ sets used here; the change must not be silently treated as a typographical correction or mixed into the original calibration. All results in this manuscript retain the original $g_{\rho N}^{2}=247.3409$ and $\Lambda_\omega=0.050$ for both sets. Thus, our comparison with the original magnetic calculation uses a specified calibration and is not a calculation with the later refit.
\begin{table}[h]
\centering
\caption{Parameters of the FSU2H and FSU2R sets, Table~I of Ref.~\cite{Tolos2017}. Masses and $\kappa$ are in MeV and $n_0$ is in $\mathrm{fm}^{-3}$.}
\begin{tabular}{lcc}
\hline
 & FSU2H & FSU2R \\
\hline
$m_N$ & 939 & 939 \\
$m_\sigma$ & 497.479 & 497.479 \\
$m_\omega$ & 782.500 & 782.500 \\
$m_\rho$ & 763.000 & 763.000 \\
$g_{\sigma N}^2$ & 102.7200 & 107.5751 \\
$g_{\omega N}^2$ & 169.5315 & 182.3949 \\
$g_{\rho N}^2$ & 247.3409 & 247.3409 \\
$\kappa$ & 4.0014 & 3.0911 \\
$\lambda$ & $-0.013298$ & $-0.001680$ \\
$\zeta$ & 0.008 & 0.024 \\
$\Lambda_\omega$ & 0.050 & 0.050 \\
$n_0$ & 0.1505 & 0.1505 \\
\hline
\end{tabular}
\label{tab:fsu2_params}
\end{table}

\subsection{Hyperon couplings}

The hyperon couplings follow the prescription of Ref.~\cite{Tolos2017}. Here the vector couplings are obtained from SU(3) flavor symmetry, vector dominance, and ideal mixing of the physical $\omega$ and $\phi$ mesons. The $\omega$ couplings are
\begin{equation}
g_{\omega\Lambda}
=
g_{\omega\Sigma}
=
\frac{2}{3}g_{\omega N},
\qquad
g_{\omega\Xi}
=
\frac{1}{3}g_{\omega N}.
\label{eq:hyperon_omega}
\end{equation}
For the $\rho$ meson,
\begin{equation}
g_{\rho\Lambda}=0,
\qquad
g_{\rho\Sigma}
=
g_{\rho\Xi}
=
g_{\rho N}.
\label{eq:hyperon_rho}
\end{equation}
These relations use the isospin convention appearing explicitly in the Lagrangian: $I_3=\pm1/2$ for nucleons and $\Xi$ hyperons, and $I_3=+1,0,-1$ for the $\Sigma$ triplet. The relative factor of two between the effective $\rho$ couplings of charged $\Sigma$ hyperons and nucleons is therefore supplied by $I_3$.

For the $\phi$ meson, we include the 20\% reduction of the $\phi$--$\Lambda$ coupling adopted in Ref.~\cite{Tolos2017}:
\begin{equation}
\begin{aligned}
g_{\phi N} &= 0,\\
g_{\phi\Lambda} &=
-0.8\,\frac{\sqrt{2}}{3}g_{\omega N},\\
g_{\phi\Sigma} &=
-\frac{\sqrt{2}}{3}g_{\omega N},\\
g_{\phi\Xi} &=
-\frac{2\sqrt{2}}{3}g_{\omega N}.
\end{aligned}
\label{eq:hyperon_phi}
\end{equation}
This reduction was introduced to reproduce the weak attraction inferred from the $\Lambda\Lambda$ bond energy.

The scalar couplings are calibrated to the hyperon potentials in symmetric nuclear matter at saturation \cite{Tolos2017}:
\begin{subequations}
\begin{eqnarray}
U_{\Lambda}^{(N)}(n_0) &=& -28~\mathrm{MeV},\\
U_{\Sigma}^{(N)}(n_0) &=& +30~\mathrm{MeV},\\
U_{\Xi}^{(N)}(n_0) &=& -18~\mathrm{MeV}.
\end{eqnarray}
\label{eq:hyperon_potentials}
\end{subequations}
In symmetric nuclear matter, the calibration condition is
\begin{equation}
U_Y^{(N)}(n_0)
=
-g_{\sigma Y}\bar{\sigma}_0
+
g_{\omega Y}\bar{\omega}_0,
\label{eq:hyperon_potential_calibration}
\end{equation}
where $\bar{\sigma}_0$ and $\bar{\omega}_0$ are the nucleonic mean fields at saturation. We adopt the published FSU2H scalar-coupling ratios of Ref.~\cite{Tolos2017}:
\begin{equation}
\begin{aligned}
g_{\sigma\Lambda} &= 0.6113\,g_{\sigma N},\\
g_{\sigma\Sigma} &= 0.4665\,g_{\sigma N},\\
g_{\sigma\Xi} &= 0.3157\,g_{\sigma N}.
\end{aligned}
\label{eq:hyperon_scalar}
\end{equation}
For FSU2R, the scalar couplings are obtained from the same potentials, Eq.~(\ref{eq:hyperon_potentials}), through Eq.~(\ref{eq:hyperon_potential_calibration}) with the FSU2R saturation fields, giving $g_{\sigma\Lambda}=0.6126\,g_{\sigma N}$, $g_{\sigma\Sigma}=0.4608\,g_{\sigma N}$, and $g_{\sigma\Xi}=0.3168\,g_{\sigma N}$. The same coupling is assigned to all charge states within each hyperon isospin multiplet. No additional hidden-strangeness scalar meson is introduced.

\subsection{Anomalous magnetic moment couplings}

The anomalous magnetic moment couplings $\kappa_b$ for all baryons are taken at their vacuum values, using Eq.~(\ref{eq:amm_definition}) and the magnetic moments listed in Table~\ref{tab:amm_values}. These values correspond to the standard phenomenological prescription employed in the neutron-star literature \cite{Broderick2000,Yue2009}. The $\Sigma^0$ magnetic moment is treated as the estimate specified in the table caption.

We note that in-medium modifications of baryon magnetic moments may be significant at the densities relevant for neutron star cores \cite{Ryu2010a,Ryu2010b}. A self-consistent treatment of density-dependent AMMs within the RMF framework remains an open problem and is deferred to future work. However, the field-dependent radiative corrections must be interpreted relative to the mass scale of the particle concerned \cite{Schwinger1998,ManrezaParet2014}. As discussed in Sec.~\ref{sec:introduction}, fields of order $10^{18}$ G remain below the critical scale set by the vacuum baryon masses, although they greatly exceed the electron critical field. Our results therefore describe AMM effects within the phenomenological vacuum-baryon-AMM framework, with possible in-medium modifications treated as a separate uncertainty.

\subsection{Magnetic-field configurations}\label{sec:Bconfig}
We consider two  configurations of the magnetic field. The first assumes a constant uniform field $\overline B$ for the matter calculation, following Broderick et al.~\cite{Broderick2000}, using a density average of the profile defined below over $n_0\leq n_B\leq6n_0$:
\begin{equation}
	\overline B
	=
	\frac{1}{5n_0}
	\int_{n_0}^{6n_0}B(n_B)\,dn_B
	=
	8.88619\times10^{17}~\mathrm{G}.
	\label{eq:average_constant_field}
\end{equation}
This is an average over baryon density, not a volume-weighted average over a stellar configuration. The same constant field is used in the calculations with and without AMM. Where the constant field is labeled approximately $10^{18}$ G, this rounded label refers to the value in Eq.~(\ref{eq:average_constant_field}).

Our second field configuration uses the density-dependent profile  from Tolos et al. \cite{Tolos2017}:
\begin{equation}
B(n_B)
=
B_s+B_c
\left[
1-\exp\left(
-\beta\left(\frac{n_B}{n_0}\right)^\gamma
\right)
\right],
\label{eq:magnetic_profile}
\end{equation}
with
\begin{equation}
\begin{aligned}
B_s &= 10^{15}~\mathrm{G},\\
B_c &= 2\times10^{18}~\mathrm{G},\\
\beta &= 0.0065,\\
\gamma &= 3.5.
\end{aligned}
\label{eq:magnetic_profile_parameters}
\end{equation}
The profile represents a phenomenological increase of the field toward the stellar interior, reaching $B\simeq10^{18}$ G at $n_B\simeq3.8n_0$ and approaching its limiting value at higher densities.

We note that both magnetic field configurations are phenomenological. In reality, the magnetic field configuration in a neutron star interior is determined self-consistently by the coupled Einstein--Maxwell equations. However, we adopt the widely used phenomenological approaches where the magnetic field is treated as a fixed background, allowing us to focus on the microphysical effects of Landau quantization and AMM on the EoS. The treatment of the crust and the continuation of the field into the outer stellar layers are specified with the stellar-structure calculation.

\subsection{Landau-level filling and numerical method}

For charged baryons, the effective Fermi energy is
\begin{equation}
E_{F,b}=\mu_b-V_b,
\end{equation}
where $V_b=g_{\omega b}\bar{\omega}+g_{\rho b}I_3^b\bar{\rho}+g_{\phi b}\bar{\phi}$. For a given spin projection $s=\pm1$, the longitudinal Fermi momentum is
\begin{equation}
k_{F,\nu,s}^{b\,2}
=
E_{F,b}^{2}
-
\left(
\sqrt{m_b^{*2}+2\nu|q_b|B}
-
s\kappa_b B
\right)^2.
\label{eq:charged_fermi_momentum}
\end{equation}
A Landau level contributes only when the spin--Landau state is allowed and $k_{F,\nu,s}^{b\,2}\geq0$. The highest occupied Landau level in the field regime considered here is therefore spin dependent:
\begin{equation}
\nu_{\max}^{b,s}
=
\left\lfloor
\frac{(E_{F,b}+s\kappa_b B)^2-m_b^{*2}}
{2|q_b|B}
\right\rfloor.
\label{eq:nu_max_spin_numerical}
\end{equation}
The occupation condition in Eq.~(\ref{eq:charged_fermi_momentum}) and the allowed combinations in Eq.~(\ref{eq:landau_spin_relation}) are imposed explicitly. For leptons, for which AMM are neglected, this reduces to
\begin{equation}
\nu_{\max}^{\ell}
=
\left\lfloor
\frac{\mu_\ell^2-m_\ell^2}
{2|q_\ell|B}
\right\rfloor.
\end{equation}

At each baryon density $n_B$, the coupled nonlinear system consisting of the meson-field equations, charge neutrality, baryon-number conservation, and $\beta$-equilibrium is solved self-consistently. The unknowns are the meson fields together with the independent chemical potentials $\mu_n$ and $\mu_e$. The solution at the previous density point is used as the initial guess for the next point, improving convergence and tracking the physical branch continuously. Landau levels are summed explicitly up to $\nu_{\max}^{i,s}$ for each charged species and allowed spin state. The zero-field calculation uses the corresponding continuum momentum integrals.

The step-like behavior in the particle fractions and magnetization originates from the discrete changes in $\nu_{\max}^{i,s}$ as density or magnetic field strength varies. These changes produce physical de Haas--van Alphen-type features of the magnetized Fermi gas. A sufficiently fine density grid and convergence checks are required to distinguish these features from numerical artifacts.

\section{RESULTS}

We have implemented the formalism described in Sec.~II and solved the coupled RMF equations numerically for the FSU2H parameter set; the FSU2R set is used in Sec.~\ref{sec:stiffening} and for the stellar sequences of Table~\ref{tab:tovmasses_fsu2r}. The code computes the particle fractions, Landau-level filling, magnetization, equation of state, and mass-radius relations. 
We compare the zero-field case with the two magnetic-field configurations described in Sec.~\ref{sec:Bconfig}. The constant-field configuration is calculated with and without baryon AMM, and the density-dependent configuration includes baryon AMM.
 All composition and EoS comparisons within a given figure or table use the same parameters and hyperon couplings; unless stated otherwise they are for FSU2H.
\subsection{Particle Fractions}

\begin{figure}[h]
\centering
\includegraphics[width=\columnwidth]{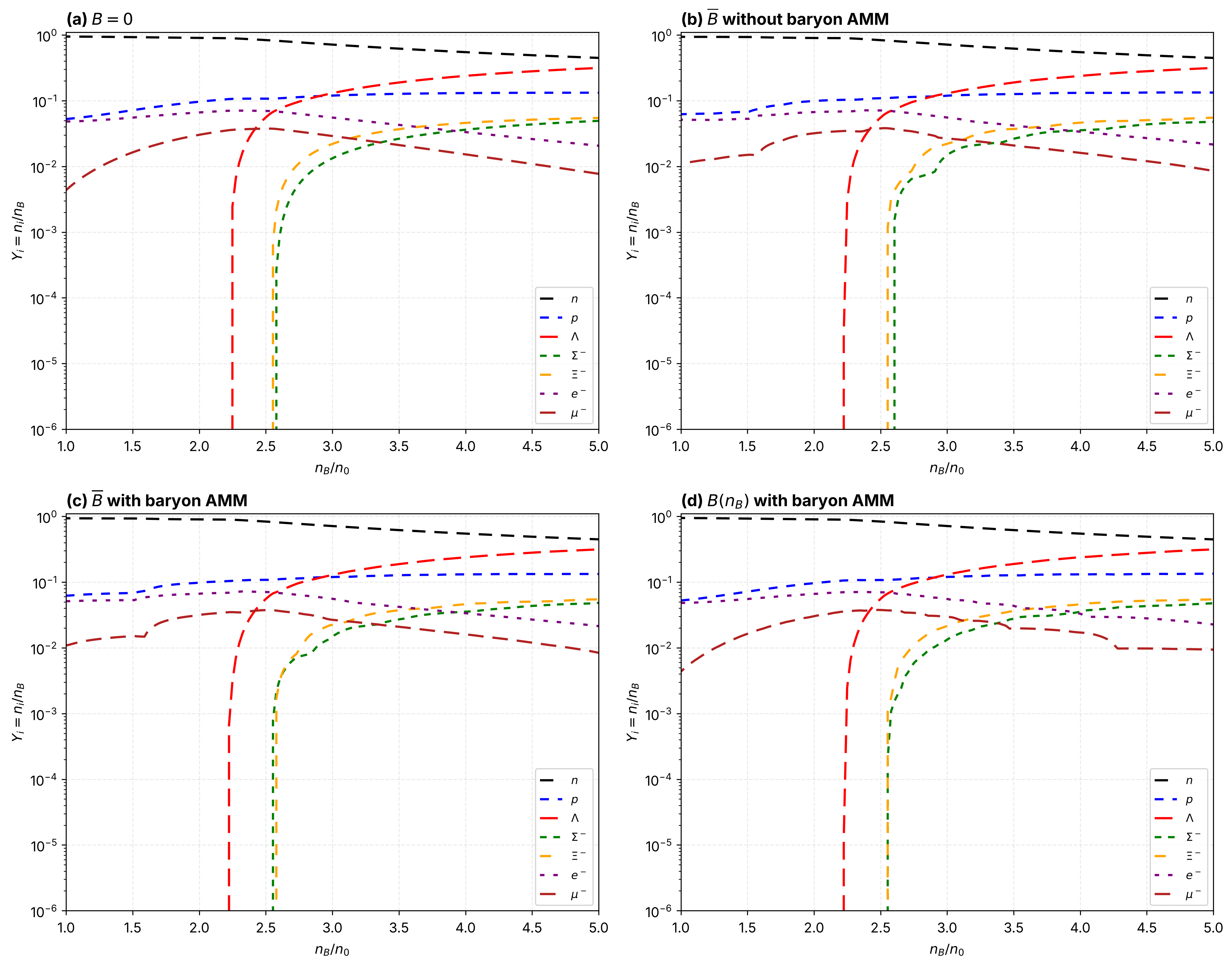}
\caption{Particle fractions $Y_i=n_i/n_B$ for charge-neutral, $\beta$-equilibrated hyperonic matter for the indicated magnetic-field configurations and AMM couplings. 
}
\label{fig:fractions}
\end{figure}

Figure~\ref{fig:fractions} presents the particle fractions $Y_i=n_i/n_B$ for four magnetic-field configurations. Panel (a) shows the zero-field case, serving as the baseline. Panel (b) shows $\overline B$ from Eq.~(\ref{eq:average_constant_field}) without AMM couplings. Panel (c) shows $\overline B$ including all baryon AMM couplings. Panel (d) shows $B(n_B)$  from Eq.~(\ref{eq:magnetic_profile}) with all baryon AMM couplings. The following subsections analyze each case in turn.

\subsubsection{${B=0}$}
In the absence of a magnetic field, the neutron fraction decreases from $Y_n\simeq0.928$ at $n_B=1.5n_0$ to $Y_n\simeq0.448$ at $5n_0$. The proton fraction rises from approximately $0.0718$ to $0.1324$. The first populated grid points place the $\Lambda$ onset near $2.24n_0$, followed by the $\Xi^{-}$ near $2.54n_0$ and the $\Sigma^{-}$ near $2.57n_0$. These onset estimates have a density-grid resolution of approximately $0.025n_0$. At $5n_0$, the fractions are $Y_\Lambda\simeq0.3153$, $Y_{\Sigma^-}\simeq0.0494$, and $Y_{\Xi^-}\simeq0.0547$.

\subsubsection{$\overline B$ without AMM coupling}

Applying the constant field magnetic $\overline{B}$ without AMM modifies the charged-particle populations and their equilibrium with the neutral baryons. At $1.5n_0$, the proton fraction is approximately $0.0681$, compared with $0.0718$ at zero field. At $5n_0$, the fractions are $Y_p\simeq0.1330$, $Y_\Lambda\simeq0.3157$, $Y_{\Sigma^-}\simeq0.0477$, and $Y_{\Xi^-}\simeq0.0552$. Thus, the magnetic response is species dependent: the $\Sigma^{-}$ fraction is reduced at this density, whereas the $\Lambda$ fraction is slightly larger than at zero field. Landau quantization does not produce a uniform suppression of all hyperons in this calculation.

\subsubsection{$\overline B$ with AMM coupling}

Including the baryon AMM modifies the composition through the spin-dependent terms in the single-particle spectra. At $5n_0$, the fractions are $Y_p\simeq0.13299$, $Y_\Lambda\simeq0.31566$, $Y_{\Sigma^-}\simeq0.04848$, and $Y_{\Xi^-}\simeq0.05465$. Relative to the same-field calculation without AMM, the $\Sigma^{-}$ population increases, while the $\Xi^{-}$ population decreases. The $\Lambda$ fraction changes only slightly. The changes are therefore smaller and more species dependent than a general restoration of the hyperon population would suggest.

For the $\Lambda$, with $\kappa_\Lambda/\mu_N\simeq-0.613$, the energetically favoured spin branch is lowered. However, the total population depends on both spin branches, the chemical potentials, and the self-consistent meson fields. Lowering one branch does not by itself determine the sign of the change in the equilibrium fraction. In-medium modifications of the AMM could further change this balance \cite{Ryu2010b}.

\subsubsection{$B(n_B)$ with AMM coupling}

For $B(n_B)$ from Eq.~(\ref{eq:magnetic_profile}) with $(\beta, \gamma )=(0.0065, 3.5)$, the field is weaker at low density and increases towards the high-density core. At $5n_0$, the fractions are $Y_p\simeq0.1346$, $Y_\Lambda\simeq0.3153$, $Y_{\Sigma^-}\simeq0.0478$, and $Y_{\Xi^-}\simeq0.0547$. The $\Lambda$ fraction remains close to its zero-field value, while the charged-particle fractions exhibit the response to the local field. For the adopted $(\beta, \gamma )$ parameters, $B(n_B)$ reaches $B\simeq10^{18}$ G at $n_B\simeq3.8n_0$. Next, we consider different $(\beta, \gamma )$ values for $B(n_B)$.

\subsection{ $\mathbf{B(n_B) }$ with different $\mathbf{(\beta, \gamma )}$ parameters}

\begin{figure}[h]
\centering
\includegraphics[width=\columnwidth]{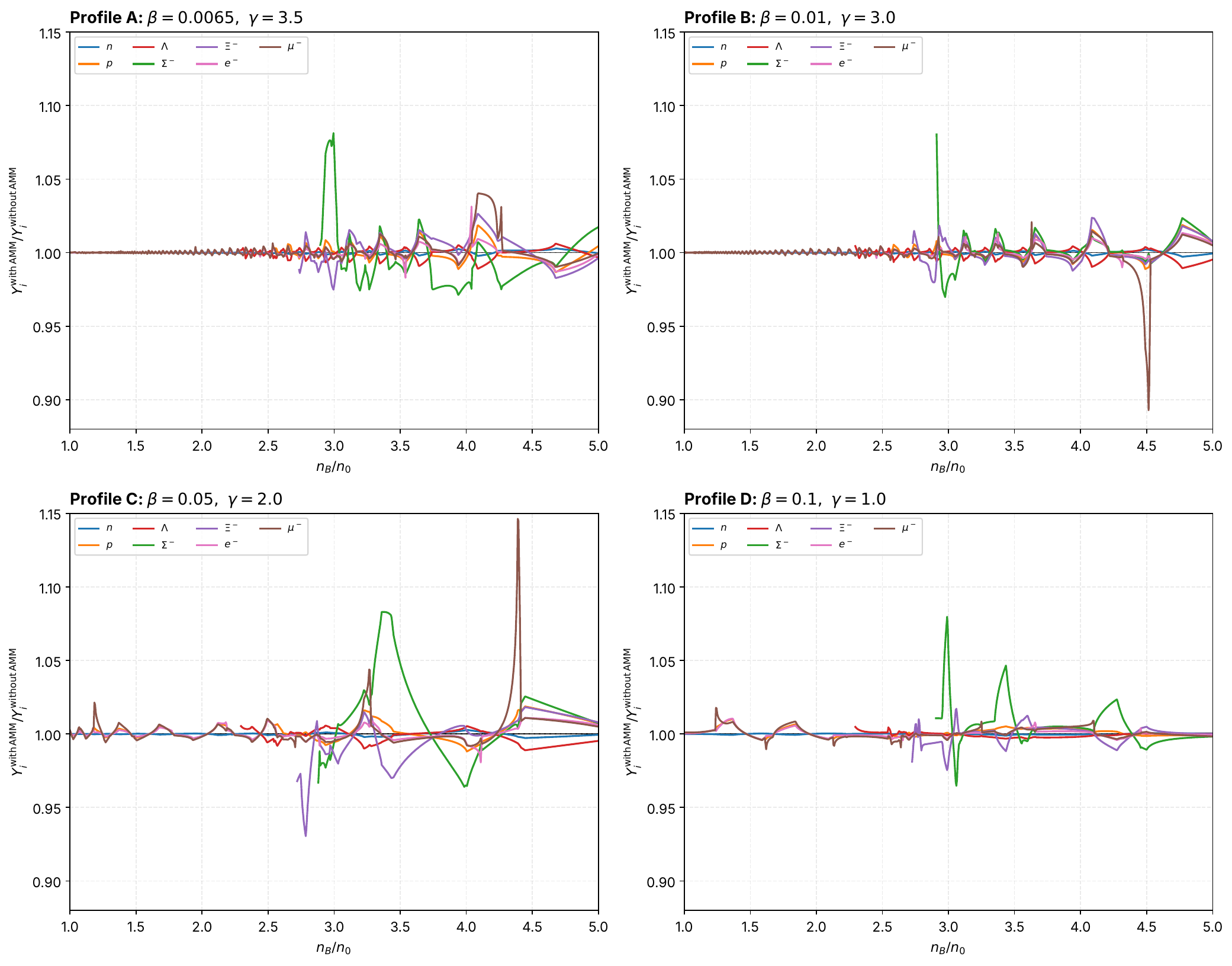}
\caption{Ratio of the particle fractions with baryon AMM to those without baryon AMM, $Y_i^{\mathrm{with\,AMM}}/Y_i^{\mathrm{without\,AMM}}$, for four density-dependent magnetic-field profiles $B(n_B)$. From top left to bottom right, $(\beta,\gamma)=(0.0065,3.5)$, $(0.01,3.0)$, $(0.05,2.0)$, and $(0.10,1.0)$. All profiles use $B_s=10^{15}$ G and $B_c=2\times10^{18}$ G. Hyperon ratios are shown where $Y_i>10^{-2}$ and the other species where $Y_i>10^{-3}$; species are identified by colour.
}
\label{fig:density_profiles}
\end{figure}

From Eq.~(\ref{eq:magnetic_profile_parameters}) the $(\beta,\gamma)$ parameters control the local field strength of $B(n_B)$ at each density and consequently the Landau-level occupation and equilibrium composition. The broad composition trends are similar among the profiles, while the charged-particle fractions and population thresholds show profile-dependent variations \cite{Tolos2017}. However, the impact of the baryon AMM couplings were not established. Figure~\ref{fig:density_profiles} shows the ratio of the particle fractions with the baryon AMM couplings to those without them, $Y_i^{\mathrm{with\,AMM}}/Y_i^{\mathrm{without\,AMM}}$, for four $(\beta,\gamma)$ parameter $B(n_B)$ profiles. The ratios are evaluated on a fine grid of $801$ densities, $\Delta n_B=0.005\,n_0$, between $n_0$ and $5n_0$. Hyperon ratios are shown where $Y_i>10^{-2}$, which excludes the steep region just above each threshold where a small shift of the onset density produces a large ratio, and the ratios of all other species where $Y_i>10^{-3}$. The muon fraction stays above $4\times10^{-3}$ at all densities, so the muon curves are present throughout.

The effect of the baryon AMM on the composition is negligible at low density and becomes appreciable at high density. Below $2n_0$ the ratios differ from unity by at most $0.4\%$ for $(\beta,\gamma)=(0.0065,3.5)$ and $(0.01,3.0)$ and by about $2\%$ for $(0.05,2.0)$ and $(0.1,1.0)$, whose fields are stronger at low density. Above about $2.5n_0$, where the fields are strong, the fractions of the charged hyperons change by up to about $8\%$ for $\Sigma^{-}$ (all profiles) and by up to about $7\%$ for $\Xi^{-}$ (reduction, $(0.05,2.0)$), while the $n$, $p$ and $\Lambda$ fractions change by less than about $2\%$. Narrow features of $10$--$15\%$ in the muon ratio at $n_B\simeq4.4$--$4.5\,n_0$ for $(0.01,3.0)$ and $(0.05,2.0)$ coincide with Landau-level thresholds. The oscillatory structure of the ratios is consistent with shifts of the Landau-level and spin-branch thresholds when the AMM are included. The result is therefore a percent-level, species-dependent modification that grows with the local field strength.

We checked the numerical consistency of these calculations on all fine-grid sweeps ($10$ sweeps, $8010$ points). In every point the self-consistent solution satisfies the residual condition with a norm below $9\times10^{-9}$, baryon number is conserved to better than $10^{-15}$, charge neutrality holds to better than $2\times10^{-9}$, and the energy density and pressure increase monotonically with density. Only $2$ of the $8010$ points required a density shift, of $10^{-5}$ in relative terms. The coupling constants are those of Table~I of Ref.~\cite{Tolos2017}, as listed in Table~\ref{tab:fsu2_params}.

\subsection{Landau Level Filling}
\begin{figure}[h]
\centering
\includegraphics[width=\columnwidth]{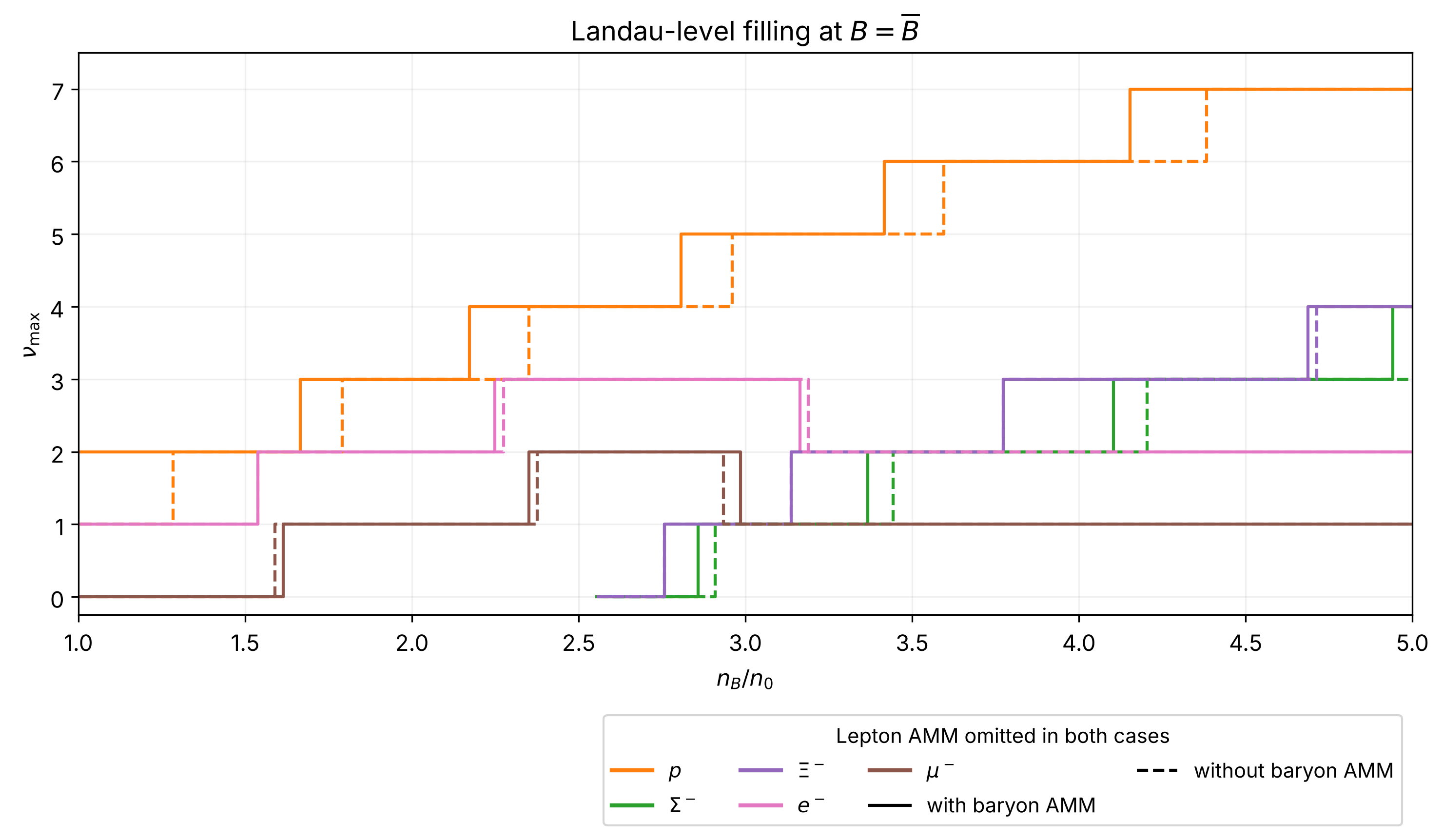}
\caption{Maximum occupied Landau-level index $\nu_{\mathrm{max}}$ for $p$, $\Sigma^-$, $\Xi^-$, $e^-$, and $\mu^-$ at $B=\overline B$. Solid and dashed lines show the equilibrium solutions with and without baryon AMM, respectively. Unoccupied species are omitted. Lepton AMM are neglected in both calculations.}
\label{fig:landau}
\end{figure}

The profile-dependent changes in the charged-particle fractions motivate a closer examination of Landau-level occupation. We use the constant comparison field to isolate the effect of baryon AMM on the filling thresholds.
Figure~\ref{fig:landau} shows the highest occupied Landau-level index, $\nu_{\mathrm{max}}$, of protons, leptons and negatively charged hyperons for magnetic field $B=\overline B$, these are selected for this diagnostic; the plot does not enumerate all charged species.  When baryon AMM couplings are included (``with baryon AMM'' in Fig.~\ref{fig:landau}) $\nu_{\mathrm{max}}$ refers to the largest occupied index among the allowed spin branches. The step-like structure reflects changes in Landau-level occupation. 
The  AMM on and AMM off curves refer only to the baryon AMM in the full equilibrium calculation, with lepton AMM omitted in both cases. Lepton filling follows Eq.~(\ref{eq:nu_max_lepton}), rather than on $\mu_e/m_e$ alone.

Three observations emerge from Fig.~\ref{fig:landau}:
\begin{enumerate}
\item Equal charge magnitudes do not imply equal Landau-level filling, because the effective masses, chemical potentials, and AMM couplings differ among species.

\item AMM shift the baryonic filling thresholds. Changes in Landau-level occupation also contribute to the oscillatory magnetization. Related nonperiodic oscillations can occur without AMM when different charged species populate Landau levels at different thresholds \cite{Scurto2024}.

\item Although electron AMM are neglected, the electron filling thresholds can change indirectly through the chemical-equilibrium response to baryon AMM.
\end{enumerate}

\subsection{Equation of State}

\begin{figure}[h]
\centering
\includegraphics[width=\columnwidth]{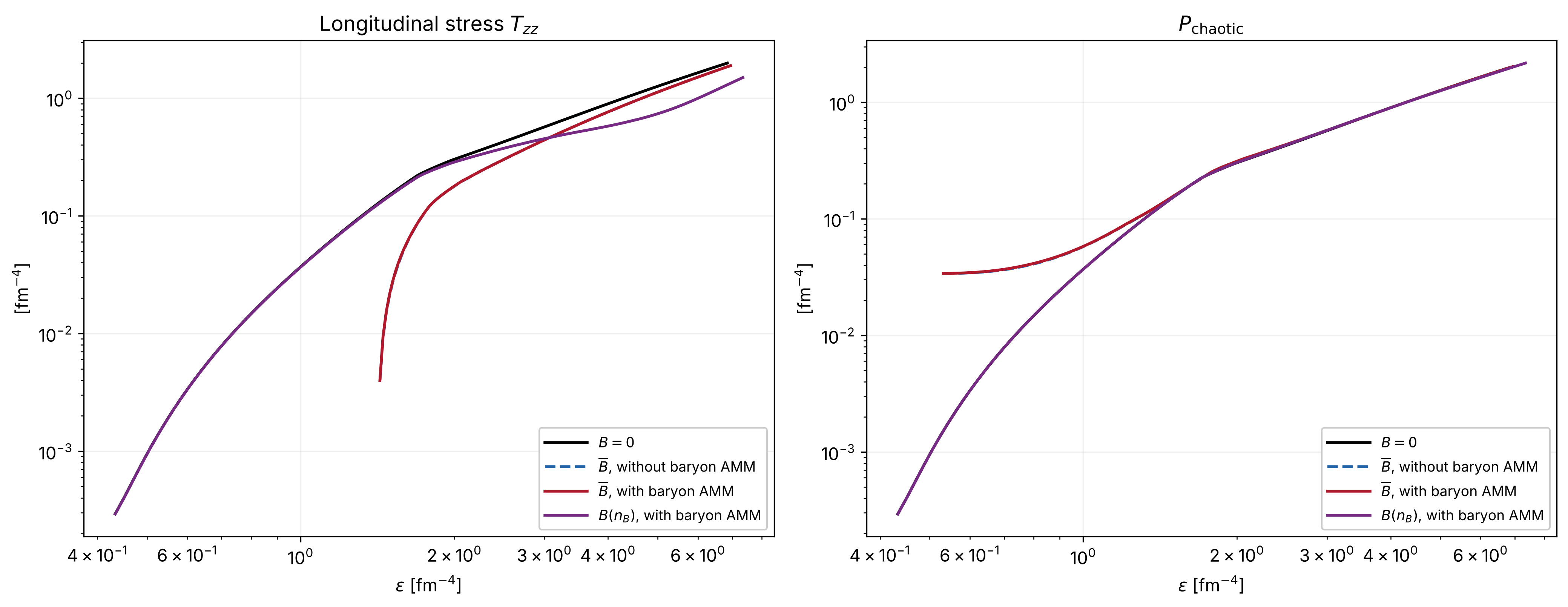}
\caption{Longitudinal stress $T_{zz}$ (left) and chaotic-field pressure $P_{\mathrm{chaotic}}$ (right) as functions of total energy density for the indicated magnetic-field configurations and baryon AMM couplings.}
\label{fig:eos}
\end{figure}

Figure~\ref{fig:eos} presents the magnetic-field dependence of the EoS. The longitudinal stress in the left panel is distinguished from the scalar thermodynamic pressure. The right panel shows the chaotic-field pressure used in the spherical stellar calculations. The plotted quantities are $T_{zz}=P_{\mathrm{matt}}-B^2/(8\pi)$ and $P_{\mathrm{chaotic}}=P_{\mathrm{matt}}+B^2/(24\pi)$, respectively. Both are plotted against total energy density, including the magnetic-field energy. The constant-field cases use $B=\overline B=8.88619\times10^{17}$ G. Nonpositive longitudinal stresses are not displayed on the logarithmic axis. The chaotic-field configuration is an isotropic stellar-structure approximation.

The AMM contribution is assessed at the same magnetic field and total energy density, using the same pressure prescription. At $B=10^{18}$ G, the chaotic-field magnetic pressure, Eq.~(\ref{eq:chaotic_pressure}),  is approximately $0.04195~\mathrm{fm}^{-4}$.
 At the comparison field $\overline B$, it is approximately $0.03313~\mathrm{fm}^{-4}$.

We define the relative change of the pressure due to the AMM, at the same field prescription and total energy density, as
\begin{equation}
\delta_{\mathrm{AMM}}(\epsilon)
=100\,\frac{P_{\mathrm{with\,AMM}}(\epsilon)-P_{\mathrm{without\,AMM}}(\epsilon)}
{P_{\mathrm{without\,AMM}}(\epsilon)}.
\label{eq:amm_relative_pressure_difference}
\end{equation}
Positive and negative values indicate stiffening and softening, respectively, at the specified energy density. Table~\ref{tab:eos_comparison} gives pressures obtained by solving at $\epsilon=4.0~\mathrm{fm}^{-4}$, rather than reading approximate values from the plotted curves.

\begin{table}[h]
\centering
\caption{Chaotic-field pressure at total energy density $\epsilon=4.0~\mathrm{fm}^{-4}$.}
\begin{tabular}{lc}
\hline
EoS sequence & $P_{\mathrm{chaotic}}$ ($\mathrm{fm}^{-4}$) \\
\hline
$B=0$ & 0.897659 \\
 $\overline{B}$, no AMM & 0.897244 \\
$\overline B$, with AMM & 0.898173 \\
$B(n_B)$, with AMM & 0.897771 \\
\hline
\end{tabular}
\label{tab:eos_comparison}
\end{table}

At this energy density, including AMM couplings increase the constant-field pressure by approximately $9.29\times10^{-4}~\mathrm{fm}^{-4}$, or $0.104\%$. The AMM-included constant-field pressure is also slightly above the zero-field and density-dependent values at this point. This ordering is not universal. At $\epsilon=3.0~\mathrm{fm}^{-4}$, 
including the AMM couplings for $\overline{B}$ the differencce is approximately $-0.0082\%$, indicating a small local softening. At $\epsilon=5.0~\mathrm{fm}^{-4}$, it is approximately $+0.0485\%$, although both the $\overline{B}$ pressures--with and without AMM--are then below the $B=0$ pressure. The effect of including the AMM couplings is therefore weak and varies with energy density; it cannot be represented by a uniform increase/decrease.

\subsection{AMM-induced stiffening and the mass--central-density relation for FSU2H and FSU2R}\label{sec:stiffening}

\begin{figure}[h]
\centering
\includegraphics[width=\columnwidth]{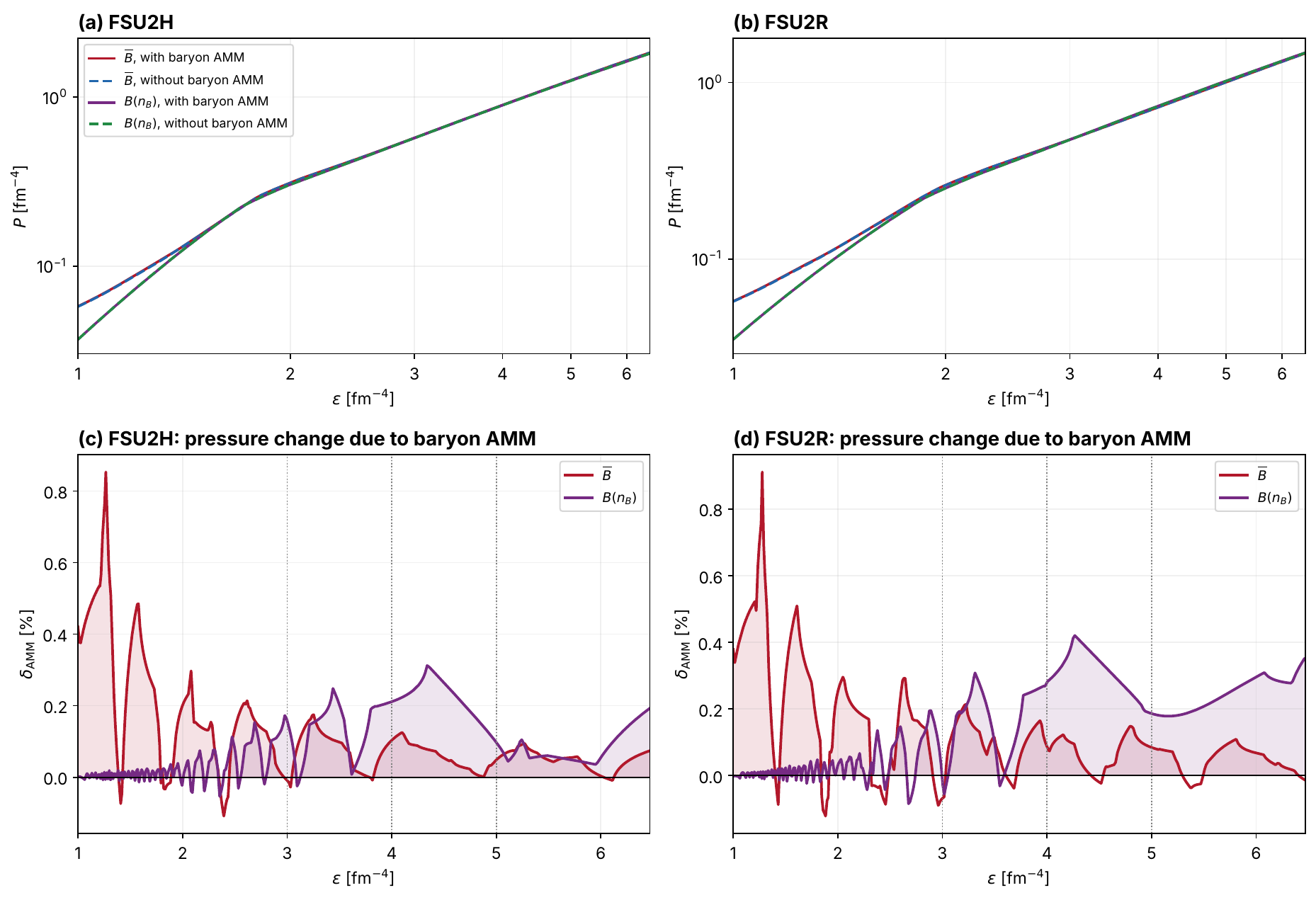}
\caption{Pressure versus total energy density for FSU2H (left) and FSU2R (right). Upper panels: $P(\epsilon)$ for $\overline B$ and $B(n_B)$, with solid (dashed) curves including (omitting) baryon AMM. Lower panels: $\delta_{\mathrm{AMM}}$, Eq.~(\ref{eq:amm_relative_pressure_difference}), for $\overline B$ and $B(n_B)$; positive values indicate that the AMM stiffen the EoS.
}
\label{fig:eos_amm}
\end{figure}

The comparison at $\epsilon=3$, $4$ and $5~\mathrm{fm}^{-4}$ above samples the effect of the AMM at three points. To establish its sign and size over the whole density range, and to test whether it depends on the parameterization, Fig.~\ref{fig:eos_amm} shows the pressure as a function of the total energy density for FSU2H and FSU2R, for the constant field $\overline B$ and for the density-dependent field $B(n_B)$, each with and without baryon AMM. The pressure is $P_{\mathrm{chaotic}}$ throughout. The upper panels show the EoS, on which the effect of the AMM is too small to be seen by eye; the lower panels therefore show $\delta_{\mathrm{AMM}}$, Eq.~(\ref{eq:amm_relative_pressure_difference}), the pressure with AMM relative to the pressure without AMM at the same field prescription and total energy density; positive values indicate stiffening.

In both parameterizations the solid curves lie above the dashed ones over most of the range. Quantitatively, $\delta_{\mathrm{AMM}}$, Eq.~(\ref{eq:amm_relative_pressure_difference}), is positive for $93\%$ of $2\leq\epsilon\lesssim6.4~\mathrm{fm}^{-4}$ for FSU2H with $\overline B$ (mean $+0.08\%$ for $2\leq\epsilon\leq6~\mathrm{fm}^{-4}$) and for $95\%$ with $B(n_B)$ (mean $+0.12\%$); for FSU2R the corresponding numbers are $84\%$ ($+0.08\%$) and $93\%$ ($+0.19\%$). The AMM therefore stiffen the EoS in both parameterizations and for both field configurations, and the effect is small: typical values are a few tenths of a percent or less, with the largest values, about $1\%$ for $\overline B$ and $0.3$--$0.4\%$ for $B(n_B)$, at the lowest energy densities shown, below the hyperon onset. The correction is not uniform in sign. Narrow windows of softening are present, with minima of $-0.11\%$ (FSU2H) and $-0.12\%$ (FSU2R) for $\overline B$, and $-0.05\%$ and $-0.09\%$ for $B(n_B)$, and the oscillatory structure is consistent with the crossing of Landau-level and spin-branch thresholds. At $\epsilon=3$, $4$ and $5~\mathrm{fm}^{-4}$, $\delta_{\mathrm{AMM}}$ is $-0.008\%$, $+0.10\%$, and $+0.05\%$ for FSU2H with $\overline B$ and $-0.068\%$, $+0.087\%$, and $+0.084\%$ for FSU2R with $\overline B$; for $B(n_B)$ it is $+0.16\%$, $+0.21\%$, $+0.10\%$ (FSU2H) and $-0.019\%$, $+0.28\%$, $+0.19\%$ (FSU2R).

\begin{figure}[h]
\centering
\includegraphics[width=\columnwidth]{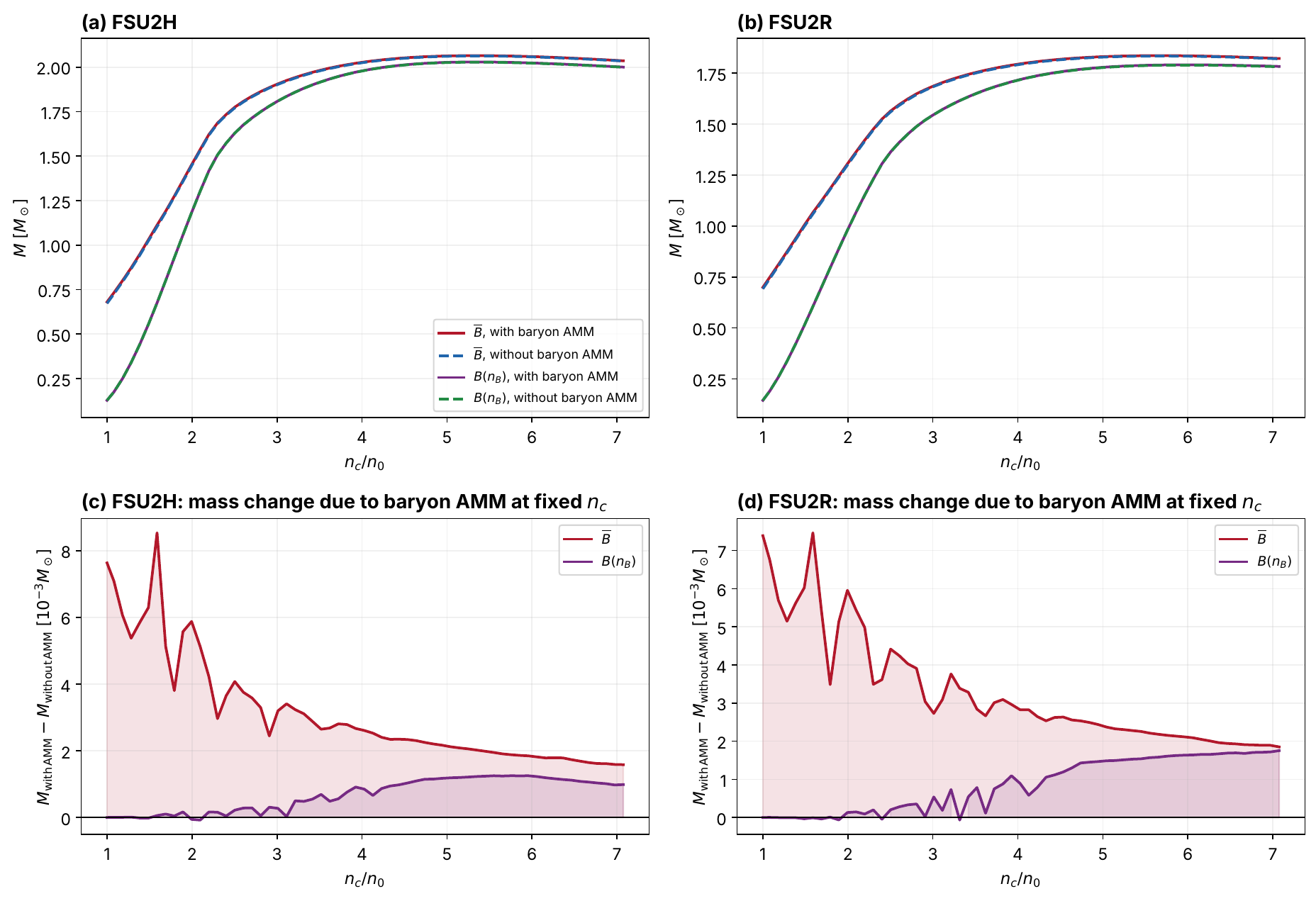}
\caption{Gravitational mass as a function of central baryon density for FSU2H (left) and FSU2R (right). Upper panels: $M(n_c)$ for $\overline B$ and $B(n_B)$, with solid (dashed) curves including (omitting) baryon AMM. Lower panels: change of the mass due to the baryon AMM at the same $n_c$, $M_{\mathrm{AMM}}-M_{\mathrm{no\,AMM}}$, for $\overline B$ and $B(n_B)$; positive values mean that the AMM increase the mass.}
\label{fig:mass_density}
\end{figure}

The consequence for the star is shown in Fig.~\ref{fig:mass_density}, which gives the gravitational mass as a function of the central baryon density $n_c$. The TOV equations are integrated at the central pressure of each tabulated core point, so that the calculations with and without AMM are compared at exactly the same $n_c$. For $\overline B$, the mass is larger with AMM at every sampled central density $n_c\geq n_0$, by $2.7\times10^{-3}M_\odot$ on average for $2\leq n_c/n_0\leq6$ (FSU2H) and $3.0\times10^{-3}M_\odot$ (FSU2R), and by up to $8.5\times10^{-3}M_\odot$ and $7.5\times10^{-3}M_\odot$, respectively. For $B(n_B)$ the mass is larger with AMM for $98\%$ (FSU2H) and $96\%$ (FSU2R) of the sampled densities $n_c\geq2n_0$, with mean gains of $0.7\times10^{-3}M_\odot$ and $0.9\times10^{-3}M_\odot$. A stiffer EoS supports more mass at the same central density, so this is the stellar counterpart of the pressure increase in Fig.~\ref{fig:eos_amm}, and it has the same sign in the two parameterizations.

\subsection{Magnetization}

\begin{figure}[h]
\centering
\includegraphics[width=\columnwidth]{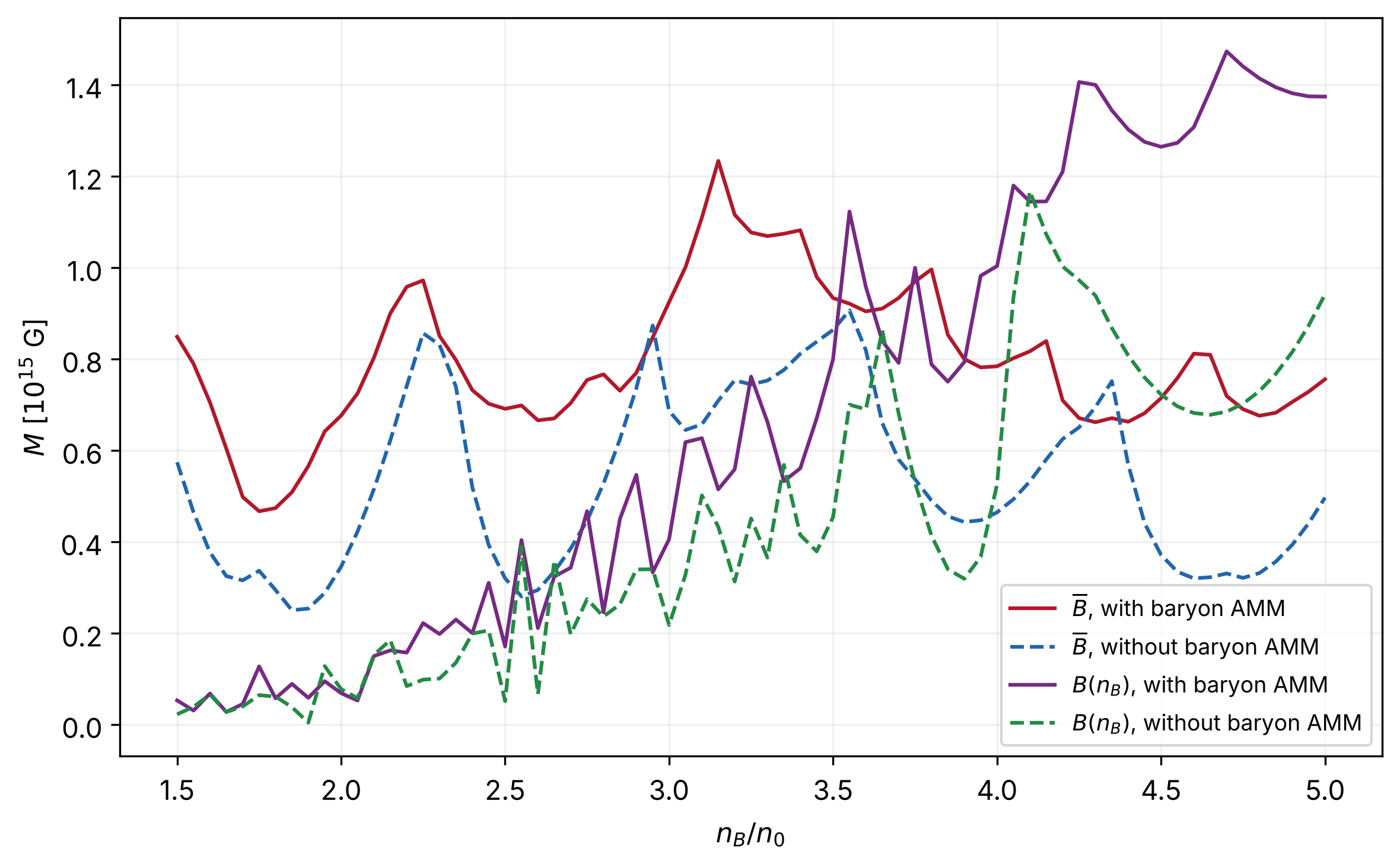}
\caption{Magnetization, Eq.~\ref{eq:magnetization_fixed_density}, as a function of $n_B$ for the constant field $\overline B$ and $B(n_B)$ with $(\beta=0.0065,\gamma=3.5)$ . Solid lines include baryon AMM couplings and dashed lines omit them. 
}
\label{fig:magnetization}
\end{figure}

Figure~\ref{fig:magnetization} shows the magnetization from Eq.~\ref{eq:magnetization_fixed_density}
for $\overline{B}$ and $B(n_B)$ with $(\beta=0.0065,\gamma=3.5)$
 magnetic field configurations, with and without baryon AMM couplings. At each point, the derivative is evaluated at fixed baryon density, with the composition and mean fields re-equilibrated. For $B(n_B)$, the derivative is evaluated around the local values and not along the profile. The field derivative is taken with respect to $B$[gauss]; resulting magnetization is expressed consistently in Gaussian units.

The magnetization is positive at the sampled densities, indicating paramagnetic behaviour. For the constant field, the AMM-included magnetization ranges from approximately $0.467$ to $1.234$ (in units of $10^{15}$ G), compared with $0.250$ to $0.907$ without AMM. Including AMM couplings generally enhance the response, but shifts in the oscillation pattern mean that the enhancement is not positive at every density. Representative values are listed in Table~\ref{tab:magnetization}.

\begin{table}[h]
\centering
\footnotesize
\caption{Representative sample of magnetization in units of $10^{15}$ G. 
}
\begin{tabular}{c|c|c|c|c}
\hline
$n_B/n_0$ & $\overline{B}$, no AMM & $\overline{B}$, with AMM & $B(n_B)$, with AMM  & $B(n_B)$, no AMM  \\
\hline
1.5 & 0.574 & 0.849 & 0.053 & 0.023 \\
2.0 & 0.346 & 0.676 & 0.069 & 0.078 \\
2.5 & 0.320 & 0.691 & 0.171 & 0.051 \\
3.0 & 0.687 & 0.925 & 0.406 & 0.218 \\
3.5 & 0.864 & 0.934 & 0.800 & 0.454 \\
4.0 & 0.465 & 0.784 & 1.004 & 0.526 \\
4.5 & 0.372 & 0.715 & 1.265 & 0.722 \\
5.0 & 0.497 & 0.756 & 1.375 & 0.942 \\
\hline
\end{tabular}
\label{tab:magnetization}
\end{table}

\subsection{Mass-radius relations and maximum masses}

To determine the mass-radius relations, we solve the Tolman--Oppenheimer--Volkoff equations,
\begin{subequations}
	\begin{eqnarray}
\frac{dP}{dr}
&=&-\frac{G}{c^2}
\frac{[\epsilon(r)+P(r)][m(r)+4\pi r^3P(r)/c^2]}
{r^2[1-2Gm(r)/(rc^2)]},
\label{eq:tov_pressure}
\\
\frac{dm}{dr}&=&\frac{4\pi r^2\epsilon(r)}{c^2}.
\label{eq:tov_mass}
\end{eqnarray}
\end{subequations}
Here $\epsilon$ is the total energy density, $P$ is the pressure, and $m(r)$ is the enclosed gravitational mass. For magnetized configurations, $P=P_{\mathrm{chaotic}}$ is used.

The core EoS is joined to the tabulated crust of Sharma et al.~\cite{Sharma2015}. All 97 crust rows are retained. A linear interpolation in $\log P$ and $\log\epsilon$ connects the crust endpoint at $n_B=0.08~\mathrm{fm}^{-3}$ to the core at $0.12~\mathrm{fm}^{-3}$. This is a composite matching prescription; chemical-potential continuity across the join is not imposed. In the constant-field stellar sequences, the field rises smoothly from zero at the core matching density to $\overline B$ at $n_0$, and remains constant above $n_0$. Thus, the uniform-matter comparison and the treatment near the stellar crust are explicitly distinguished.
\begin{figure}[h]
	\centering
	\includegraphics[width=\columnwidth]{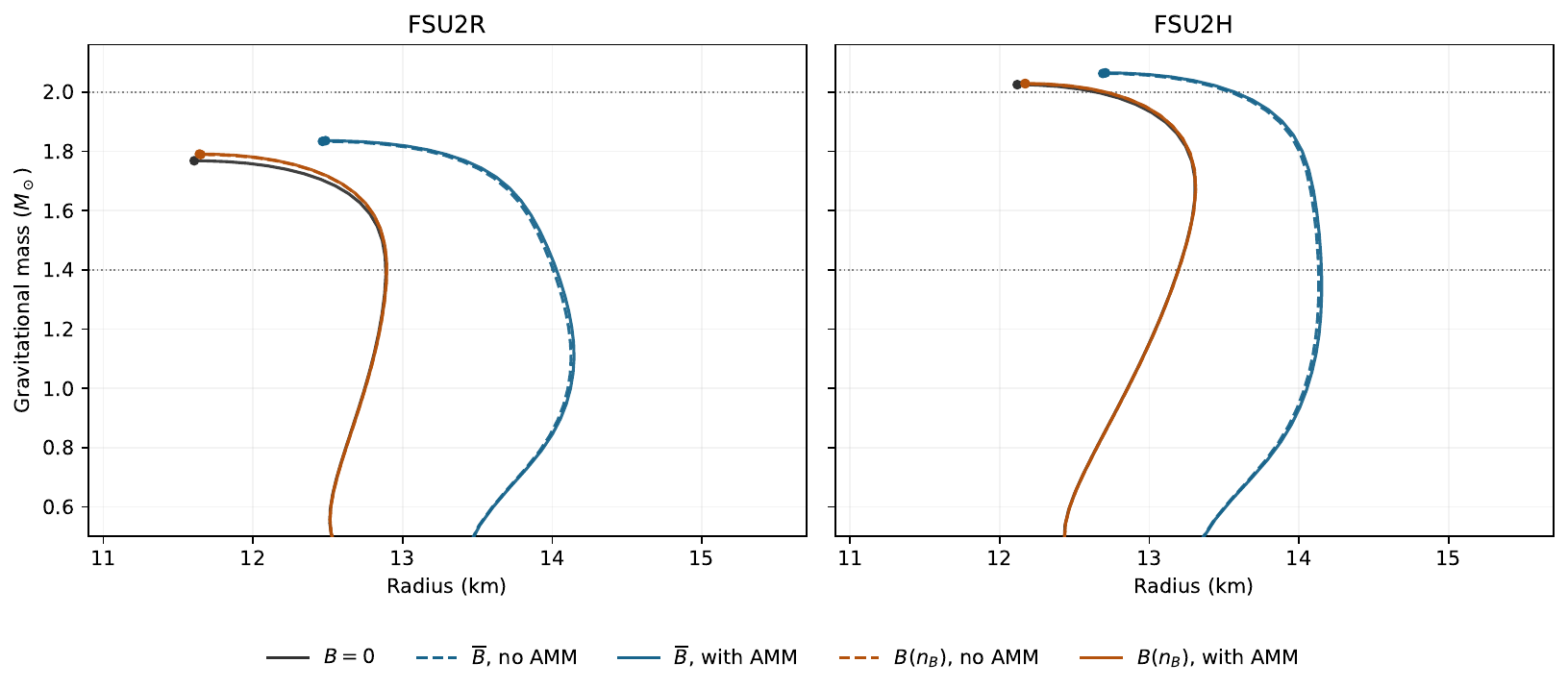}
	\caption{Mass--radius relations for FSU2R (left) and FSU2H (right), using the same tabulated crust in all ten sequences. Curves stop at the first maximum mass (dots). For both prescribed fields, solid and dashed lines correspond to calculations with and without baryon AMM. Horizontal dotted lines indicate $1.4M_\odot$ and $2M_\odot$.}
	\label{fig:massradius}
\end{figure}

The integration proceeds from a chosen central pressure to the lowest tabulated crust pressure, $9.5393\times10^{18}~\mathrm{erg\,cm}^{-3}$, which defines the numerical surface. The radius is $R$ and the gravitational mass is $M=m(R)$. Since the TOV equations assume spherical symmetry and isotropic pressure, the mass-radius curves are estimates within the chaotic-field and crust-matching approximations. Figure~\ref{fig:massradius} and Tables~\ref{tab:tovmasses_fsu2r} and~\ref{tab:tovmasses} show all five configurations for each model. At fixed $\overline B$, baryon AMM increase $M_{\max}$ by about $0.002M_\odot$ in each model. The corresponding increases in $R_{1.4}$ are $0.022$ km for FSU2R and $0.019$ km for FSU2H.

\begin{table}[h]
\centering
\footnotesize
\caption{Maximum masses and radii from the TOV sequences for FSU2R. Radii are in kilometres.}
\begin{tabular}{lccc}
\hline
EoS sequence & $M_{\max}/M_\odot$ & $R(M_{\max})$ & $R_{1.4}$ \\
\hline
$B=0$ & 1.76828 & 11.61 & 12.889 \\
$\overline{B}$, no AMM & 1.83373 & 12.47 & 14.008 \\
$\overline{B}$, with AMM & 1.83591 & 12.49 & 14.029 \\
$B(n_B)$, no AMM & 1.78919 & 11.65 & 12.895 \\
$B(n_B)$, with AMM & 1.79082 & 11.64 & 12.895 \\
\hline
\end{tabular}
\label{tab:tovmasses_fsu2r}
\end{table}

\begin{table}[h]
\centering
\footnotesize
\caption{Maximum masses and radii from the TOV sequences for FSU2H. Radii are in kilometres.}
\begin{tabular}{lccc}
\hline
EoS sequence & $M_{\max}/M_\odot$ & $R(M_{\max})$ & $R_{1.4}$ \\
\hline
$B=0$ & 2.02510 & 12.12 & 13.194 \\
$\overline{B}$, no AMM & 2.06296 & 12.69 & 14.131 \\
$\overline{B}$, with AMM & 2.06498 & 12.71 & 14.150 \\
$B(n_B)$, no AMM & 2.02804 & 12.17 & 13.196 \\
$B(n_B)$, with AMM & 2.02927 & 12.17 & 13.196 \\
\hline
\end{tabular}
\label{tab:tovmasses}
\end{table}

Refining the TOV integration changes the maximum masses by less than $8\times10^{-6}M_\odot$, while halving the density-grid resolution changes them by less than $6.4\times10^{-5}M_\odot$. These checks address numerical convergence within the adopted implementation, separately from its agreement with published stellar benchmarks.

\section{DISCUSSION}

The comparison between the constant and density-dependent magnetic field configurations adopted here reveals differences in the composition, pressure, and stellar structure of hyperonic matter. In the uniform-matter calculation, $\overline{B}$ is fixed by Eq.~(\ref{eq:average_constant_field}); in the stellar calculation it tapers near the crust. For $B(n_B)$, Eq.~(\ref{eq:magnetic_profile}), field grows with density and reaches $B\simeq10^{18}$ G near $3.8n_0$. Equal density-averaged fields do not imply equal spatial field distributions or equal effects on a stellar sequence.

Direct comparison of the resultant $\overline{B}$ EoS, also considering the baryon AMM couplings, show a small AMM-induced stiffening at $\epsilon=4~\mathrm{fm}^{-4}$: the $\overline{B}$ pressure increases by approximately $0.104\%$ relative to the calculation not including the AMM couplings. However, the correction changes with energy density and includes small intervals of softening, as illustrated by the negative correction at $\epsilon=3~\mathrm{fm}^{-4}$. At $\epsilon=5~\mathrm{fm}^{-4}$, including the AMM couplings again stiffen the EoS relative to the calculation without AMM, while the pressure remains below the $B=0$ value. Thus, stiffening relative to the magnetized no AMM case must be distinguished from stiffening relative to zero magnetic field. Figures~\ref{fig:eos_amm} and \ref{fig:mass_density} show that the stiffening relative to the no-AMM calculation is present over most of the energy-density range, and the corresponding mass gain at fixed central density is present over almost all of the density range, for both FSU2H and FSU2R and for both $\overline B$ and $B(n_B)$, so it is not specific to the parameter set or to the field configuration.

At $1.4M_\odot$, introducing the constant core field without AMM increases the radius by $1.119$ km for FSU2R and $0.937$ km for FSU2H. We checked the pressure and energy density of all 97 crust points in each of the ten sequences and used the same matching densities and surface pressure. To isolate the explicit field contribution, we retained the magnetized matter EoS and field taper but omitted both $B^2/(8\pi)$ from the energy density and $B^2/(24\pi)$ from the pressure in a diagnostic TOV integration. The FSU2R radius then falls from $14.008$ to $12.906$ km, close to its zero-field value of $12.889$ km. For FSU2H the corresponding radii are $14.131$, $13.207$, and $13.194$ km. The radius increase therefore arises predominantly from the explicit field terms in this prescribed core field, which reaches full strength at $n_0$. The density-dependent field is weaker in the outer core and produces a much smaller radius shift.

The constant magnetic field  is useful as a diagnostic of magnetic effects \cite{Broderick2000}. It permits a direct comparison of the Landau-quantized EoS with and without AMM. The density-dependent magnetic field configuration, $B(n_B)$ provides a phenomenological description of a field increasing towards the stellar centre. Neither configuration determines the magnetic geometry self-consistently as that would require a coupled Einstein--Maxwell calculation \cite{Sinha2026}.

The particle fractions show the combined effects of Landau quantization, spin splitting, and the chemical-equilibrium conditions. The response is species dependent. At $5n_0$, adding AMM to the constant-field calculation increases the $\Sigma^{-}$ fraction and reduces the $\Xi^{-}$ fraction, while changing the $\Lambda$ fraction only slightly. The updated calculation therefore does not support a general description in which Landau quantization strongly suppresses all hyperons and AMM restore them towards their zero-field populations.

Changes in hyperon fractions alone do not determine the sign of the pressure correction. The pressure also depends on the spin-dependent spectra and on the feedback through the RMF mean fields. The small changes in composition must therefore be interpreted together with the direct comparison at fixed total energy density. The maximum-mass increase of approximately $0.002M_\odot$ between the constant-field AMM-off and AMM-on sequences is consistent with a weak structural effect. A pressure difference at one energy density, by itself, is not sufficient to predict the change in maximum mass.

The result depends on the magnetic-moment prescription. In-medium changes to nucleon and hyperon moments can alter their relative magnetic response and hence the equilibrium composition \cite{Ryu2010a,Ryu2010b}. Our calculation uses vacuum baryon moments and does not introduce density-dependent AMM. Its quantitative conclusions apply to that prescription and the adopted FSU2H calibration.

The applicability of the linear AMM coupling must be assessed relative to the mass scale of the particle concerned. As discussed in Sec.~\ref{sec:introduction}, the fields considered here remain below the critical scale set by the vacuum baryon masses. The strong-field radiative results discussed by Schwinger~\cite{Schwinger1998} and Manreza Paret et al.~\cite{ManrezaParet2014} do not by themselves establish a suppression of the phenomenological baryon AMM used here. Electrons are in a different field regime, and their weak-field AMM correction should not be extrapolated linearly to these fields. Possible modifications of baryon magnetic moments in dense matter remain a separate uncertainty.

Landau-level filling and changes in the charged-particle fractions may affect weak-interaction rates. Maruyama et al.~\cite{Maruyama2014} studied asymmetric neutrino emission in magnetized proto-neutron-star matter. Our composition calculations provide possible inputs to such studies, but no neutrino emissivities or transport coefficients are calculated here. Quantitative conclusions about cooling or neutrino-emission asymmetries require a separate transport calculation.

The magnetization displays oscillations in both the with AMM and no AMM cases. AMM generally enhance the magnetic response, although shifts in the oscillatory structure prevent a pointwise enhancement at every density. Nonperiodic oscillations are not unique to AMM: Landau quantization of multiple charged species can also produce them \cite{Scurto2024}. 

The pressure treatment remains a limitation. Scalar thermodynamic pressure must be distinguished from directional stress components, and magnetization-current forces must be included in mechanical equilibrium \cite{Blandford1982,Potekhin2012,Chatterjee2015}. The chaotic-field prescription averages the Maxwell stress for use in spherical stellar calculations. It does not replace a calculation of an ordered magnetic field and the associated stellar deformation. The composite crust join and the taper of the constant core field are additional modelling choices that affect the stellar radii.

\subsection{Comparison with previous magnetized RMF calculations}

The present calculation follows the treatment of Landau quantization and baryon AMM developed by Broderick et al.~\cite{Broderick2000,Broderick2002}. Their work established that these effects can modify the composition and EoS of strongly magnetized matter. Here the comparison is performed within the original FSU2H parameterization of Tolos et al.~\cite{Tolos2017}, retaining the printed scalar hyperon coupling ratios and the prescribed vector couplings.

The zero-field hyperonic sequences give maximum masses of $1.768M_\odot$ (FSU2R) and $2.025M_\odot$ (FSU2H), consistent with the rounded values $1.77M_\odot$ and $2.03M_\odot$ in Ref.~\cite{Tolos2017}. We adopt the RMF parameter values in their Table~I and their density-dependent field profile. Our magnetic comparisons use pairs of sequences computed here with the same prescribed field and with baryon AMM switched on or off.

A related input check concerns the hyperon potentials. Keeping the printed scalar ratios $0.6113$, $0.4665$, and $0.3157$ gives saturation potentials of approximately $-27.51$, $+27.84$, and $-17.60$ MeV for $\Lambda$, $\Sigma$, and $\Xi$, respectively, in the present implementation. These differ slightly from the calibration targets $-28$, $+30$, and $-18$ MeV. The ratios have been retained rather than refitted, so agreement with both the printed couplings and the target potentials is not asserted.

The calculation also complements the chaotic-field study of Wu et al.~\cite{Wu2017} and the magnetized hyperonic calculations of Rather et al.~\cite{Rather2021}. These studies use different model choices, so the magnitude and sign of composition changes need not agree point by point. The sensitivity of hyperonic RMF predictions to their couplings, also examined by Sanson et al.~\cite{Sanson2026}, reinforces the need to interpret the present results as a model-specific calculation.

\section{CONCLUSIONS}

Within the FSU2H and FSU2R models, baryon AMM stiffen the magnetized hyperonic EoS over most of the energy-density range, for both the constant field $\overline B$ and the density-dependent field $B(n_B)$. The effect is small, typically about $0.1\%$ in pressure at fixed total energy density, and is accompanied by narrow windows of softening, for example at $\epsilon=3~\mathrm{fm}^{-4}$ for $\overline B$. The AMM-included pressure is therefore not uniformly higher than the no-AMM, zero-field, or density-dependent-field results. AMM also modify the composition and magnetization and increase the maximum mass by approximately $0.002M_\odot$ for $\overline B$ in both parameterizations within the adopted spherical approximation. At fixed central baryon density the mass is larger with AMM at every sampled density for $\overline B$ and at $96$--$98\%$ of the sampled densities ($n_c\geq2n_0$) for $B(n_B)$.

\end{document}